\documentclass{fmj2010}
\usepackage{graphicx}

\def\fermi{\textit{Fermi}}

\def\lesssim{\lower4pt\hbox{${\buildrel < \over \sim}$}}
\def\gtrsim{\lower4pt\hbox{${\buildrel > \over \sim}$}}
\begin{document}
   \title{Models for the Spectral Energy Distributions and
   Variability of Blazars}

   \author{M. B\"ottcher\inst{1}
          }

   \institute{Astrophysical Institute, Department of Physics and Astronomy,
   Ohio University, Athens, OH, USA}

\abstract{
In this review, recent progress in theoretical models for blazar emission
will be summarized. The salient features of both leptonic and lepto-hadronic
approaches to modeling blazar spectral energy distributions will be reviewed.
I will present sample modeling results of spectral energy distributions (SEDs)
of different types of blazars along the blazar sequence, including \fermi\
high-energy $\gamma$-ray data, using both types of models. Special emphasis
will be placed on the implications of the recent very-high-energy (VHE)
$\gamma$-ray detections of non-traditional VHE $\gamma$-ray blazars,
including intermediate and low-frequency-peaked BL Lac objects and
even flat-spectrum radio quasars. Due to the featureless optical spectra
of BL Lac objects, the redshifts of several BL Lacs remain unknown. I
will briefly discuss possible constraints on their redshift using
spectral modeling of their SED including \fermi\ + ground-based VHE
$\gamma$-ray data. It will be shown that in some cases,
spectral modeling with time-independent single-zone models alone is not
sufficient to constrain models, as both leptonic and lepto-hadronic models
are able to provide acceptable fits to the overall SED. Subsequently,
recent developments of time-dependent and inhomogeneous blazar models
will be discussed, including detailed numerical simulations as well
as a semi-analytical approach to the time-dependent radiation signatures
of shock-in-jet models.
}

   \maketitle
%

\section{\label{intro}Introduction}

\begin{figure*}
\centering
\includegraphics[width=0.95\textwidth]{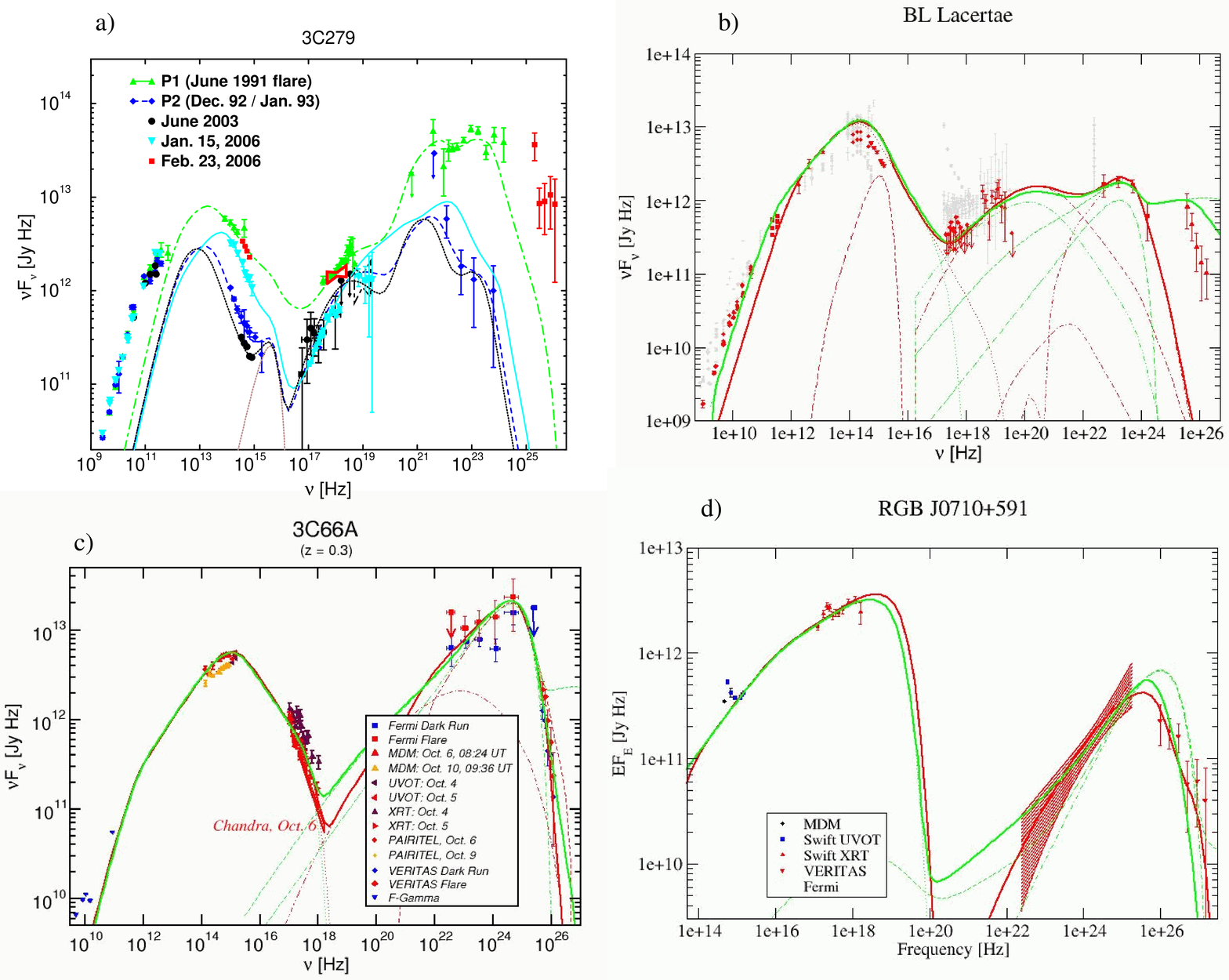}
\caption{\label{SEDs}Spectral energy distributions of four sub-classes
of blazars: a) the FSRQ 3C279 (from \cite{collmar10}), b) the LBL BL Lacertae,
(data from \cite{abdo10b}), c) the intermediate BL Lac 3C66A (data from
\cite{acciari10c}), and d) the HBL RGB J0710+591 (data from \cite{acciari10b}).
In Panel a) (3C279), lines are one-zone leptonic model fits to SEDs at
various epochs shown in the figure. In all other panels, red lines
are fits with a leptonic one-zone model; green lines are fits
with a one-zone lepto-hadronic model. }
\end{figure*}

Blazars (BL~Lac objects and $\gamma$-ray loud flat spectrum
radio quasars [FSRQs]) are the most extreme class of active
galaxies known. They have been observed at all wavelengths,
from radio through VHE $\gamma$-rays. The broadband continuum
SEDs of blazars are dominated by non-thermal emission and consist
of two distinct, broad components: A low-energy component
from radio through UV or X-rays, and a high-energy component
from X-rays to $\gamma$-rays (see, e.g., Figure \ref{SEDs}).

Blazars are sub-divided into several types, defined by the location
of the peak of the low-energy (synchrotron) SED component. Low-synchrotron-peaked
(LSP) blazars, consisting of flat-spectrum radio quasars and low-frequency
peaked BL Lac objects (LBLs), have their synchrotron peak in the infrared
regime, at $\nu_s \le 10^{14}$~Hz. Intermediate-synchrotron-peaked (ISP)
blazars, consisting of LBLs and intermediate BL Lac objects (IBLs) have
their synchrotron peak at optical -- UV frequencies at $10^{14} \, {\rm Hz}
< \nu_s \le 10^{15}$~Hz, while High-synchrotron-peaked (HSP) blazars,
almost all known to be high-frequency-peaked BL Lac objects (HBL), have
their synchrotron peak at X-ray energies with $\nu_s > 10^{15}$~Hz
(\cite{abdo10b}). This sequence had first been identified by \cite{fossati98},
and associated also with a trend of overall decreasing bolometric luminosity
as well as decreasing $\gamma$-ray dominance along the sequence FSRQ $\to$
LBL $\to$ HBL. According to this classification, the bolometric power output
of FSRQs is known to be strongly $\gamma$-ray dominated, in particular during
flaring states, while HBLs are expected to be always synchrotron dominated.
However, while the overall bolometric-luminosity trend still seems to hold,
recently, even HBLs seem to undergo episodes of strong $\gamma$-ray dominance.
A prominent example was the recent joint \fermi\ + HESS observational campaign
on the HBL PKS~2155-304 (\cite{aharonian09}).

Figure \ref{SEDs} shows examples of blazar SEDs along the blazar
sequence, from the FSRQ 3C279 (a), via the LBL BL~Lacertae (b) and
the IBL 3C~66A (c), to the HBL RGB J0710+591 (d). The sequence of increasing
synchrotron peak frequency is clearly visible. However, the \fermi\
spectrum of the LBL BL~Lacertae indicates a $\gamma$-ray flux clearly
below the synchrotron level, while the SED of the IBL 3C~66A is clearly
dominated by the \fermi\ $\gamma$-ray flux, in contradiction with the
traditional blazar sequence.

The emission from blazars is known to be variable at all wavelengths.
In particular the high-energy emission from blazars can easily
vary by more than an order of magnitude between different observing
epochs (\cite{vmont95,muk97,muk99}). However, high-energy variability
has been observed on much shorter time scales, in some cases
even down to just a few minutes (\cite{aharonian07,albert07a}).
The flux variability of blazars is often accompanied by spectral
changes. Typically, the variability amplitudes are the largest and
variability time scales are the shortest at the high-frequency
ends of the two SED components. In HBLs, this refers to
the X-ray and VHE $\gamma$-ray regimes. Such differential
spectral variability is sometimes associated with inter-band
or intra-band time lags as well as variability patterns which can
be characterized as spectral hysteresis in hardness-intensity diagrams
(e.g.,\cite{takahashi96,kataoka00,fossati00,zhang02}). However,
even within the same object this feature tends not to be persistent
over multiple observations. Also in other types of blazars, hints
of time lags between different observing bands are occasionally
found in individual observing campaigns (e.g., \cite{boettcheral07,horan09}),
but the search for time-lag patterns persisting throughout multiple
years has so far remained unsuccessful (see, e.g., \cite{hartman01} for a
systematic search for time lags between optical, X-ray and $\gamma$-ray
emission in the quasar 3C279).

\section{\label{models}Basic features of leptonic and lepto-hadronic models}

The high inferred bolometric luminosities, rapid variability,
and apparent superluminal motions provide compelling evidence
that the nonthermal continuum emission of blazars is produced
in $\lesssim$~1 light day sized emission regions, propagating
relativistically with velocity $\beta_{\Gamma}$c along a jet
directed at a small angle $\theta_{\rm obs}$ with respect to
our line of sight (for details on the arguments for relativistic
Doppler boosting, see \cite{schlickeiser96}). Let $\Gamma =
(1 - \beta_{\Gamma}^2)^{-1/2}$ be the bulk Lorentz factor of
the emission region, then Doppler boosting is determined by
the Doppler factor $D = (\Gamma [1 - \beta_{\Gamma} \cos\theta_{\rm obs}])^{-1}$.
Let primes denote quantities in the co-moving frame of the emission
region, then the observed frequency $\nu_{\rm obs}$ is related to
the emitted frequency through $\nu_{\rm obs} = D \, \nu' / (1 + z)$,
where $z$ is the redshift of the source, and the energy fluxes are
connected through $F_{\nu_{\rm obs}}^{\rm obs} = D^3 \, F'_{\nu'}$.
Intrinsic variability on a co-moving time scale $t'_{\rm var}$ will
be observed on a time scale $t_{\rm var}^{\rm obs} = t'_{\rm var} \,
(1 + z) / D$. Using the latter transformation along with causality
arguments, any observed variability leads to an upper limit on the
size scale of the emission region through $R \lesssim c \, t_{\rm var}^{\rm obs}
\, D / (1 + z)$.

While the electron-synchrotron origin of the low-frequency emission is
well established, there are two fundamentally different approaches
concerning the high-energy emission. If protons are not accelerated
to sufficiently high energies to reach the threshold for $p\gamma$ pion
production on synchrotron and/or external photons and to contribute
significantly to high-energy emission through proton-synchrotron
radiation, the high-energy radiationwill be dominated by emission from
ultrarelativistic electrons and/or pairs (leptonic models). In the
opposite case, the high-energy emission will be dominated by cascades
initiated by $p\gamma$ pair and pion production as well as proton,
$\pi^{\pm}$, and $\mu^{\pm}$ synchrotron radiation, while primary
leptons are still responsible for the low-frequency synchrotron
emission (lepto-hadronic models). The following sub-sections provide
a brief overview of the main radiation physics aspects of both leptonic
and lepto-hadronic models.

\subsection{\label{leptonic}Leptonic models}

In leptonic models, the high-energy emission is produced via Compton
upscattering of soft photons off the same ultrarelativistic electrons
which are producing the synchrotron emission.
Both the synchrotron photons produced within the jet (the SSC process:
\cite{mg85,maraschi92,bm96}), and external photons (the EC process)
can serve as target photons for Compton scattering. Possible sources
of external seed photons include the accretion disk radiation
(e.g., \cite{dermer92,ds93}), reprocessed optical -- UV emission
from circumnuclear material (e.g., the BLR; \cite{sikora94,bl95,gm96,dermer97}),
infrared emission from a dust torus (\cite{blazejowski00}), or
synchrotron emission from other (faster/slower) regions of the
jet itself (\cite{gk03,gt08}).

The relativistic Doppler boosting discussed above allows one to
choose model parameters in a way that the $\gamma\gamma$ absorption
opacity of the emission region is low throughout most of the high-energy
spectrum (i.e., low compactness). However, at the highest photon energies,
this effect may make a non-negligible contribution to the formation of
the emerging spectrum (\cite{aharonian08}) and re-process some of 
the radiated power to
lower frequencies. Also the deceleration of the 
jets may have a
significant impact on the observable properties of blazar 
emission
through the radiative interaction of emission regions with 
different
speed (\cite{gk03,ghisellini05}) and a varying Doppler 
factor
(\cite{bp09}). Varying Doppler factors may also be a result 
of
a slight change in the jet orientation without a substantial 
change
in speed, e.g., in a helical-jet configuration (e.g.,
\cite{vr99}). In the case of ordered magnetic-field structures
in the emission region, such a helical configuration should have
observable synchrotron polarization signatures, such as the
prominent polarization-angle swing recently observed in conjunction
with an optical + \fermi\ $\gamma$-ray flare of 3C~279 (\cite{abdo10a}).

In the most simplistic approaches, the underlying lepton (electrons
and/or positrons) distribution is ad-hoc pre-specified, either as
a single or broken power-law with a low- and high-energy cut-off.
While leptonic models under this assumption have been successful
in modeling blazar SEDs (e.g., \cite{ghisellini98}), they lack a
self-consistent basis for the shape of the electron distribution.
A more realistic approach consists of the self-consistent steady-state
solution of the Fokker-Planck equation including a physically motivated
(e.g., from shock-acceleration theory) acceleration of particles and
all relevant radiative and adiabatic cooling mechanisms (e.g.,
\cite{gt09,acciari09b,weidinger10}). The model described in \cite{acciari09b},
based on the time-dependent model of \cite{bc02}, has been used to
produce the leptonic model fits shown in Fig. \ref{SEDs}.

In order to reproduce not only broadband SEDs, but also variability
patterns, the time-dependent electron dynamics and radiation transfer
problem has to be solved self-consistently. Such time-dependent SSC
models have been developed by, e.g., \cite{mk97,kataoka00,lk00,sokolov04}.
External radiation fields have been included in such treatments in, e.g.,
\cite{sikora01,bc02,sm05}.

Leptonic models have generally been very successfully applied to model
the SEDs and spectral variability of blazars. The radiative cooling
time scales (in the observers's frame) of synchrotron-emitting electrons
in a typical $B \sim 1$~G magnetic field are of order of several hours
-- $\sim 1$~d at optical frequencies and $\lesssim 1$~hr in X-rays
and hence compatible with the observed intra-day variability. However,
the recent observation of extremely rapid VHE $\gamma$-ray variability
on time scales of a few minutes poses severe problems to simple one-zone
leptonic emission models. Even with large bulk Lorentz factors of
$\sim 50$, causality requires a size of the emitting region that might
be smaller than the Schwarzschild radius of the central black hole of
the AGN (\cite{begelman08}). As a possible solution, it has been
suggested (\cite{tg08}) that the $\gamma$-ray emission region may, in
fact, be only a small spine of ultrarelativistic plasma within a larger,
slower-moving jet. Such fast-moving small-scale jets could plausibly
be powered by magnetic reconnection in a Poynting-flux dominated jet,
as proposed by \cite{giannios09}.

\subsection{\label{hadronic}Lepto-hadronic models}

If a significant fraction of the jet power is converted into the
acceleration of relativistic protons in a strongly magnetized
environment, reaching the threshold for $p\gamma$ pion production,
synchrotron-supported pair cascades will develop (\cite{mb92,mannheim93}).
The acceleration of protons to the necessary ultrarelativistic energies
($E_p^{\rm max} \gtrsim 10^{19}$~eV) requires high magnetic fields of
several tens of Gauss to constrain the Larmor radius $R_L = 3.3 \times
10^{15} \, B_1^{-1} \, E_{19}$~cm, where $B = 10 \, B_1$~G, and $E_p
= 10^{19} \, E_{19}$~eV, to be smaller than the size of the
emission region, typically inferred to be $R \lesssim 10^{16}$~cm from
the observed variability time scale. In the presence of such high magnetic
fields, the synchrotron radiation of the primary protons (\cite{aharonian00,mp00})
and of secondary muons and mesons (\cite{rm98,mp00,mp01,muecke03})
must be taken into account in order to construct a self-consistent
synchrotron-proton blazar (SPB) model. Electromagnetic cascades can be
initiated by photons from $\pi^0$-decay (``$\pi^0$ cascade''), electrons
from the $\pi^\pm\to \mu^\pm\to e^\pm$ decay (``$\pi^\pm$ cascade''),
$p$-synchrotron photons (``$p$-synchrotron cascade''), and $\mu$-, $\pi$-
and $K$-synchrotron photons (``$\mu^\pm$-synchrotron cascade'').

M\"ucke \& Protheroe (2001) and M\"ucke et al. (2003) have shown that
the ``$\pi^0$ cascades'' and ``$\pi^\pm$ cascades'' from ultra-high
energy protons generate featureless $\gamma$-ray spectra, in contrast
to ``$p$-synchrotron cascades'' and``$\mu^\pm$-synchrotron cascades''
that produce a two-component $\gamma$-ray spectrum. In general, direct
proton and $\mu^{\pm}$ synchrotron radiation is mainly responsible for
the high energy bump in blazars, whereas the low energy bump is dominanted
by synchrotron radiation from the primary $e^-$, with a contribution from
secondary electrons.

Hadronic blazar models have so far been very difficult to investigate in a
time-dependent way because of the very time-consuming nature of the required
Monte-Carlo cascade simulations. In general, it appears that it is difficult
to reconcile very rapid high-energy variability with the radiative cooling
time scales of protons, e.g., due to synchrotron emission, which is
$t_{\rm sy}^{\rm obs} = 4.5 \times 10^5 \, (1 + z) \, D_1^{-1} B_1^{-2}
\, E_{19}^{-1}$~s (\cite{aharonian00}), i.e., of the order of several
days for $\sim 10$~G magnetic fields and typical Doppler factors
$D = 10 \, D_1$. However, rapid variability on time scales shorter
than the proton cooling time scale may be caused by geometrical effects.

In order to avoid time-consuming Monte-Carlo simulations of the hadronic
processes and cascades involved in lepto-hadronic models, one may
consider a simplified prescription of the hadronic processes.
\cite{ka08} have produced analytic fit functions to Monte-Carlo
generated results of hadronic interactions using the SOFIA code
(\cite{muecke00}). Those fits describe the spectra of the final
decay products, such as electrons, positrons, neutrinos, and
photons from $\pi^0$ decay. The use of those functions requires
a prior knowledge of the target photon field (in the SPB model
the electron-synchrotron photon field) as well as the proton
spectrum. Once the first-generation products are evaluated, one
still needs to take into account the effect of cascading, as the
synchrotron emission from most of the electrons (and positrons)
as well as $\pi^0$ decay $\gamma$-rays are produced at $\gg$~TeV
energies, where the emission region is highly opaque to $\gamma\gamma$
pair production. A quasi-analytical description of the cascades can
be found as follows.

We start with the injection rates of first-generation high-energy
$\gamma$-rays, $\dot N_{\epsilon}^0$, and pairs, $Q_e (\gamma)$,
from the analytical fit functions of \cite{ka08}. The cascades are
usually well described as linear cascades so that the optical depth
for $\gamma\gamma$ absorption, $\tau_{\gamma\gamma} (\epsilon)$
can be pre-calculated from the low-energy radiation field. Under
these conditions, the spectrum of escaping (observable) photons can
be calculated as

\begin{equation}
\dot N_{\epsilon}^{\rm esc} = \dot N_{\epsilon}^{\rm em} \, \left(
{1 - e^{-\tau_{\gamma\gamma} [\epsilon]} \over \tau_{\gamma\gamma} [\epsilon]}
\right)
\label{Ndotescape}
\end{equation}
where $\dot N_{\epsilon}^{\rm em}$ has contributions from the first-generation
high-energy photon spectrum and synchrotron emission from secondaries,
$\dot N_{\epsilon}^{\rm em} = \dot N_{\epsilon}^0 + \dot N_{\epsilon}^{\rm sy}$.
A fairly accurate evaluation of the synchrotron spectrum is achieved
with a single-electron emissivity of the form
$j_{\nu} \propto \nu^{1/3} \, e^{-\epsilon/\epsilon_0}$ with $\epsilon_0 =
b \, \gamma^2$, where $b \equiv B / B_{\rm crit}$ and $B_{\rm crit} = 4.4
\times 10^{13}$~G. The electron distribution, $N_e (\gamma)$ will be the
solution to the isotropicFokker-Planck equation in equilibrium ($\partial
\langle . \rangle / \partial t= 0$):

\begin{equation}
{\partial \over \partial \gamma} \left( \dot\gamma \, N_e [\gamma] \right)
= Q_e (\gamma) + \dot N_e^{\gamma\gamma} (\gamma) + \dot N_e (\gamma)^{\rm esc}.
\label{fp}
\end{equation}
For the high magnetic fields of $B \gtrsim 10$~G, as required for hadronically
dominated $\gamma$-ray emission, electron cooling will be dominated by
synchrotron losses, i.e.,$\dot\gamma = - \nu_0 \gamma^2$ with $\nu_0
= (4/3) \, c \, \sigma_T \, u_B / (m_e c^2)$.
We are only interested in high-energy particles emitting synchrotron radiation at
least at X-ray energies. For those particles, the synchrotron cooling time
scale is expected to be much shorter than the escape time scale so that
the escape term in Eq. \ref{fp} may be neglected. $\dot N_e^{\gamma\gamma}
(\gamma)$ in Eq. \ref{fp} is the rate of particle injection due to
$\gamma\gamma$ absorption, to be evaluated self-consistently with the radiation
field. In the $\gamma\gamma$ absorption of a high-energy photon of energy
$\epsilon$, one of the produced particles will assume the major fraction,
$f_{\gamma}$ of the photon energy. Hence, an electron/positron pair with
energies $\gamma_1 = f_{\gamma} \, \epsilon$ and $\gamma_2 = (1 - f_{\gamma})
\, \epsilon$ is produced. Furthermore realizing that every non-escaping photon
(according to Eq.\ref{Ndotescape}) will produce an electron/positron
pair, we can write the pair production rate as

\begin{equation}
\dot N_e^{\gamma\gamma} (\gamma) = f_{\rm abs} (\epsilon_1) \,
\left( \dot N_{\epsilon_1}^0 + \dot N_{\epsilon_1}^{\rm sy} \right)
+ f_{\rm abs} (\epsilon_2) \, \left( \dot N_{\epsilon_2}^0
+ \dot N_{\epsilon_2}^{\rm sy} \right)
\label{Ndotgamma}
\end{equation}
where $\epsilon_1 = \gamma/f_{\gamma}$, $\epsilon_2 = \gamma/(1 - f_{\gamma})$
and $f_{\rm abs} (\epsilon) \equiv 1 - (1 - e^{-\tau_{\gamma\gamma} [\epsilon]})
/ \tau_{\gamma\gamma} [\epsilon]$. With this approximation, we find an
implicit solution to Equation \ref{fp}:

\begin{equation}
N_e (\gamma) = {1 \over \nu_0 \gamma^2} \int\limits_{\gamma}^{\infty} d\tilde\gamma
\left\lbrace Q_e (\tilde\gamma) + \dot N_e^{\gamma\gamma} (\tilde\gamma)
\right\rbrace
\label{Nsolution}
\end{equation}
The solution (\ref{Nsolution}) is implicit in the sense that the particle
spectrum $N_e (\gamma)$ occurs on both sides of the equation as $\dot
N_e^{\gamma\gamma}$ depends on the synchrotron emissivity, which requires
knowledge of $N_e (\tilde\gamma)$, where pairs at energies of $\tilde\gamma_1
= \sqrt{\gamma / (f_{\gamma} b)}$ and $\tilde\gamma_2 = \sqrt{\gamma /
([1 - f_{\gamma}] \, b)}$ provide the majority of the radiative output
relevant for pair production at energy $\gamma$. However, for practical
applications, one may use the fact that generally, $\gamma$, the argument
on the l.h.s., is much smaller than $\tilde\gamma_{1,2}$. This condition
is fulfilled if there is no pion-induced pair injection at energies above
$\gamma_{\rm crit} = 4.4 \times 10^{13} \, ([1 - f_{\gamma}] \, B_G)^{-1}$
or $E_{\rm e, crit} = 2.3 \times 10^{19} \, ([1 - f_{\gamma}] \, B_G)^{-1}$~eV.
In a usual synchrotron-proton-blazar model setup, no substantial pair injection
above $E_{\rm e, crit}$ is expected. Therefore, Eq. \ref{Nsolution} may be
evaluated progressively, starting at the highest pair energies for which
$Q_0 (\gamma) \ne 0$ or $\dot N_{\epsilon_{1,2}}^0\ne 0$, and then using
the solution for $N_e (\gamma)$ for large $\gamma$as one progresses towards
lower values of $\gamma$.

Once the equilibrium pair distribution $N_e (\gamma)$ is known, it can be
used to evaluate the synchrotron emissivity and hence, using Eq.
\ref{Ndotescape} the observable photon spectrum.

Example model fits to several blazar SEDs using a simplified lepto-hadronic
model based on the \cite{ka08} fit functions and the above cascade
description are shown in Fig \ref{SEDs}, b) -- d) and compared to
leptonic models of the same SEDs. As the low-frequency component is
electron-synchrotron emission from primary electrons, it is not surprising
that virtually identical fits to the synchrotron component can be provided
in both types of models. In the high-frequency component, strongly peaked
spectral shapes, as, e.g., in 3C~66A and RGB J0710+591 require a strong
proton-synchrotron dominance with the cascading of higher-energy ($\gg$~TeV)
emission only making a minor contribution to the high-energy emission.
This, in fact, makes it difficult to achieve a substantial extension of
the escaping high-energy emission into the $> 100$~GeV VHE $\gamma$-ray
regime. In objects with a smoother high-energy SED, e.g., BL Lacertae
in Fig. \ref{SEDs}b, a substantially larger contribution from cascade
emission (and leptonic SSC emission) is allowed to account for a
relatively high level of hard X-ray / soft $\gamma$-ray emission. This
also allows for a substantial extension of the $\gamma$-ray spectrum
into the VHE regime.

\section{\label{fits}Spectral fits along the blazar sequence}

In the framework of leptonic models, the blazar sequence FSRQ $\to$
LBL $\to$ IBL $\to$ HBL is often modeled through a decreasing contribution
of external radiation fields to radiative cooling of electrons and
production of high-energy emission (\cite{ghisellini98}). In this sense,
HBLs have traditionally been well represented by pure synchrotron-self-Compton
models, while FSRQs often require a substantial EC component. This
interpretation is consistent with the observed strong emission lines
in FSRQs, which are absent in BL~Lac objects. At the same time, the
denser circumnuclear environment in quasars might also lead to a
higher accretion rate and hence a more powerful jet, consistent with
the overall trend of bolometric luminosities along the blazar sequence.
This may even be related to an evolutionary sequence from FSRQs to HBLs
governed by the gradual depletion of the circumnuclear environment
(\cite{bd02}).

However, in this interpretation, it would be expected that mostly
HBLs (and maybe IBLs) should be detectable as emitters of VHE $\gamma$-rays
since in LBLs and FSRQs, electrons are not expected to reach $\sim$~TeV
energies. This appears to contradict the recent VHE $\gamma$-ray detections
of lower-frequency peaked objects such as W~Comae (\cite{acciari08}), 3C66A
(\cite{acciari09a}), PKS~1424+240 (\cite{acciari10a}), BL~Lacertae
(\cite{albert07b}), S5~0716+714 (\cite{anderhub09}, and even the FSRQs
3C~279 (\cite{albert08}) and PKS~1510-089 (\cite{wb10}).

\begin{figure}
\centering
\includegraphics[width=0.48\textwidth]{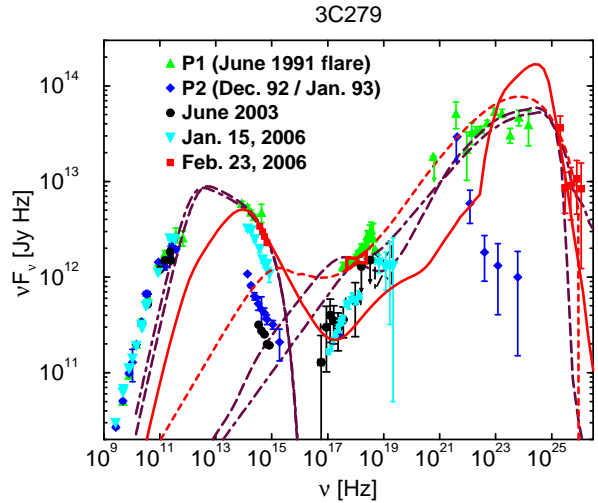}
\caption{\label{3C279MAGIC}Fits to the SED of 3C279 during the time of the
MAGIC detection (\cite{albert08}). The heavy solid curve is an attempt of a
leptonic model with EC-dominated $\gamma$-ray emission, while the dotted
curve is an SSC fit to the X-ray through $\gamma$-ray spectrum. The dashed
and dot-dashed curves are hadronic model fits. }
\end{figure}

The overall SEDs of IBLs detected by VERITAS could still be fit
satisfactorily with a purely leptonic model.  Fitting the SEDs of
the IBLs 3C66A and W~Comae with a pure SSC model, while formally
possible, would require rather extreme parameters. In particular,
magnetic fields several orders of magnitude below equipartition
would be needed, which might pose a severe problem for jet collimation.
Much more natural fit parameters can be adopted when including an EC
component with an infrared radiation field as target photons
(\cite{acciari09b,acciari10c}).

In \cite{boettcher09} it has been demonstrated that the VHE $\gamma$-ray detection
of the FSRQ 3C~279 poses severe problems for any variation of single-zone
leptonic jet model. The SED with the MAGIC points is shown in Fig. \ref{3C279MAGIC},
along with attempted leptonic as well as hadronic model fits. The fundamental
problem for the leptonic model interpretation is the extremely wide
separation between the synchrotron and $\gamma$-ray peak frequencies.
In a single-zone SSC interpretation this would require a very high
Lorentz factor of electrons at the peak of the electron distribution
and, in turn, an extremely low magnetic field. In an EC interpretation
(the solid curve in Fig. \ref{3C279MAGIC}) the SSC component, usually
dominating the X-ray emission in leptonic fits to FSRQs like 3C~279,
is too far suppressed to model the simultaneous RXTE spectrum. As an
alternative, the optical emission could be produced in a separate
emission component. A pure SSC fit to the X-ray and $\gamma$-ray
component (the short-dashed curve in Fig. \ref{3C279MAGIC}) is
technically possible, but also requires a far sub-equipartition
magnetic field. Much more natural parameters could be achieved in
a fit with the synchrotron-proton-blazar model (\cite{muecke03}),
as illustrated by the long-dashed and dot-dashed model curves in
Fig. \ref{3C279MAGIC}.

\section{\label{redshift}Redshift constraints from blazar SED modeling}

BL~Lac objects are defined by the absence of broad emission lines in
their optical spectra. This makes the determination of their redshifts,
and hence distances, very difficult, and in many cases even impossible.
Photometric redshifts, estimated under the assumption that the host
galaxies of BL Lac objects are nearly standard candles (e.g., \cite{sbarufatti05}),
are still highly uncertain, and optical spectroscopy often yields only
lower limits on redshifts (e.g., \cite{finke07}).

An alternative method of estimating BL~Lac redshifts exploits the
fact that high-energy and VHE $\gamma$-rays are absorbed through
$\gamma\gamma$ pair production on the Extragalactic Background Light
(EBL), see, e.g., \cite{dk05,stecker06,franceschini08,gilmore09,finke10}.
Substantial progress has been made in the last few years so that the
accepted models of the EBL spectrum are now converging to an EBL level
near the lower limit set by direct galaxy counts (e.g., \cite{aharonian06,abdo10c}).

Exploiting this improved knowledge of the EBL spectrum and intensity,
limits on VHE $\gamma$-ray blazars can now be set from the expectation
that the intrinsic (unabsorbed by the EBL) HE to VHE $\gamma$-ray
spectra of blazars are not becoming harder towards higher energies.
A robust upper limit to the source redshift can therefore be found
by considering a straight power-law extrapolation of the best-fit
\fermi\ power-law spectrum into the VHE $\gamma$-ray regime, and
attributing the change in spectral slope towards the measured VHE
spectrum purely to EBL absorption (e.g., \cite{georganopoulos10,acciari10a}). A
somewhat more realistic approach has been suggested by \cite{prandini10},
taking into account the expected intrinsic softening of the VHE
$\gamma$-ray spectrum, based on a sample of blazars with known
redshifts detected both by \fermi\ and ground-based VHE $\gamma$-ray
observatories.

\begin{figure}
\centering
\includegraphics[width=0.48\textwidth]{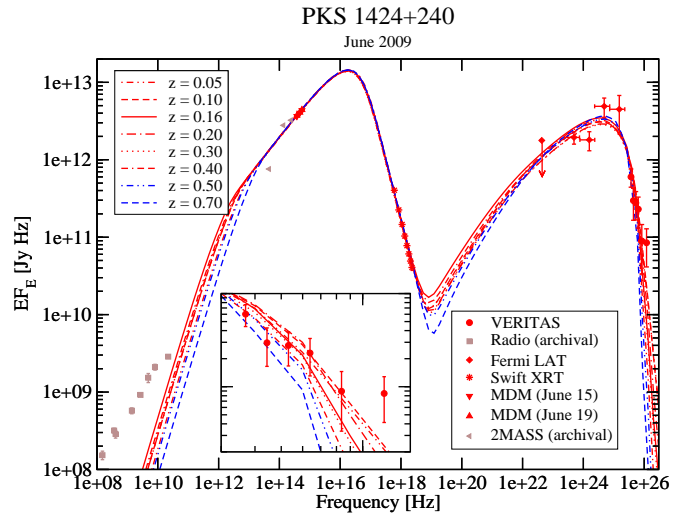}
\caption{\label{PKS1424}Fits to the SED of PKS 1424+240, assuming a
range of plausible redshifts (from \cite{acciari10a}). All SEDs have
been corrected for $\gamma\gamma$ absorption by the EBL using the model
of \cite{gilmore09}. The inset shows a zoom-in on the VHE $\gamma$-ray
spectrum. The spectral fits favour a source redshift of $z < 0.4$.}
\end{figure}

If the SED of a blazar with unknown redshift is well sampled,
the EBL absorption effect can be exploited in an even more
sophisticated way by constraining models through the SED from
the optical to GeV $\gamma$-rays and varying the redshift to
yield the best fit to the measured VHE $\gamma$-ray spectrum.
This procedure has been carried out in detail for the recent
VERITAS detections of PKS~1424+240 (\cite{acciari10a}) and
3C66A (\cite{acciari10c}).

In the case of PKS~1424+240, only very unreliable redshift
estimates were available. Fig. \ref{PKS1424} shows our suite
of models for a wide range of plausible redshifts. While a
straight power-law extrapolation of the \fermi\ spectrum yields
an upper limit of $z < 0.6$, our spectral model (leptonic SSC)
clearly favors a redshift of $z < 0.4$.
The redshift of 3C~66A is usually quoted as $z = 0.444$, but
is actually highly uncertain as well (\cite{bramel05}).
A scan through redshifts yielded a preferred redshift of
$z \sim 0.2$ -- $0.3$ for this object, in good agreement
with the estimate found by \cite{prandini10}. The leptonic
model fit to 3C66A shown in Fig. \ref{SEDs}c is based on a
redshift of $z = 0.3$.

\section{\label{inhomogeneous}Inhomogeneous jet models}

The complicated and often inconsistent variability features found
in blazars provide a strong motivation to investigate jet models
beyond a simple, spherical, one-zone geometry. The idea behind
phenomenological multi-zone models like the spine-sheath model
of \cite{tg08} or the decelerating-jet model of \cite{gk03} was
that differential relativistic motion between various emission
zones will lead to Doppler boosting of one zone's emission into
the rest frame of another zone. This can reduce the requirements
of extreme bulk Lorentz (and Doppler) factors inferred from simple
one-zone letponic modeling of rapidly variable VHE $\gamma$-ray
blazars, such as Mrk~501 or PKS~2155-304, and has led to successful
model fits to the SEDs of those sources with much more reasonable
model parameters.

However, the models mentioned above do not treat the radiation
transport and electron dynamics in a time-dependent way and do
therefore not make any robust predictions concerning variability
and inter-band cross-correlations and time lags. In order to
address those issues, much work has recently been devoted to the
investigation of the radiative and timing signatures of shock-in-jet
models, which will be summarized in the following sub-section.

\subsection{\label{shock}Shock-in-jet models}

Early versions of shock-in-jet models were developed with focus
on explaining radio spectra of extragalactic jets, e.g., by
\cite{mg85}. Their application to high-energy spectra of blazars
was proposed by \cite{spada01}. Detailed treatments of the electron
energization and dynamics and the radiation transfer in a standing
shock
(Mach disk) in a blazar jet were developed by
\cite{sokolov04,sm05,graff08}. The internal-shock model
discussed in \cite{mimica04,joshi09} assumes that the central engine is 
intermittently ejecting shells
of relativistic plasma at varying 
speeds, which subsequently collide.
Such models have had remarkable 
success in explaining SEDs and time
lag features of generic blazars.

\begin{figure}
\centering
\includegraphics[width=0.48\textwidth]{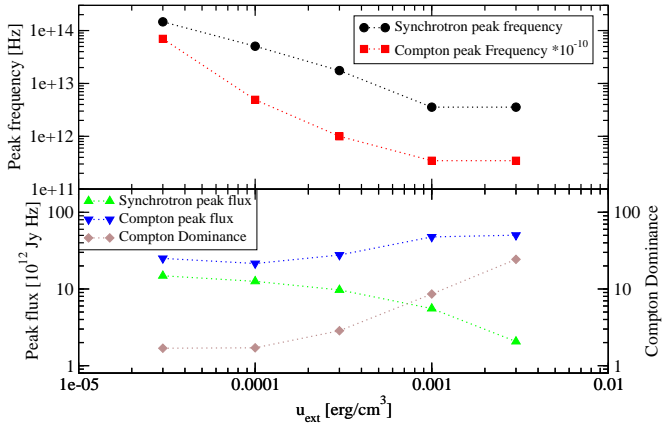}
\caption{\label{cs_sppar} Dependence of the SED characteristics of
time-averaged spectra from the intermal-shock model, on the external
radiation energy density $u_{\rm ext}$. }
\end{figure}

The realistic treatment of radiation transfer in an internal-shock
model for a blazar requires the time-dependent evaluation of retarded
radiation fields originating from all parts of the shocked regions
of the jet. The model system is therefore highly non-linear and can
generally only be solved using numerical simulations
(e.g., \cite{sokolov04,mimica04,graff08}). As the current
detailed internal-shock models employ either full expressions or
accurate approximations to the full emissivities of synchrotron
and Compton emission, a complete simulation of the time-dependent
spectra and light curves istime-consuming and does therefore
generally not allow to efficiently explore a large parameter
space. General patterns of the SED, light curves and expected
time lags between different wavelength bands have been demonstrated
for very specific, but observationally very poorly constrained,
sets of parameters.

To remedy this aspect, \cite{bd10} have developed a semi-analytical
internal-shock model for blazars. In this model, the time- and space-dependent
electron spectra, affected by shock acceleration behind the forward and
reverse shocks, and subsequent radiative cooling, is calculated fully
analytically. Taking into account all light travel time effects, the
observed synchrotron and external-Compton spectra are also evaluated
fully analytically, using a $\delta$-function approximation to the
emissivities. The evaluation of the SSC emission still requires a
two-dimensional numerical integration.

This semi-analytical model allowed the authors to efficiently scan a
substantial region of parameter space and discuss the dependence on
the characteristics of time-averaged SEDs, as well as cross-band
correlations and time lags. As an example, Fig. \ref{cs_sppar} shows
the dependence of the SED characteristics on the external radiation
field energy density, $u_{\rm ext}$. In the classical interpretation
of the blazar sequence, an increasing $u_{\rm ext}$ corresponds to
a transition from BL Lac spectral characteristics to FSRQ-like
characteristics. The decreasing synchrotron peak frequency and
increasing Compton dominance found in the parameter study are
reproducing this effect.

\begin{figure}
\centering
\includegraphics[width=0.48\textwidth]{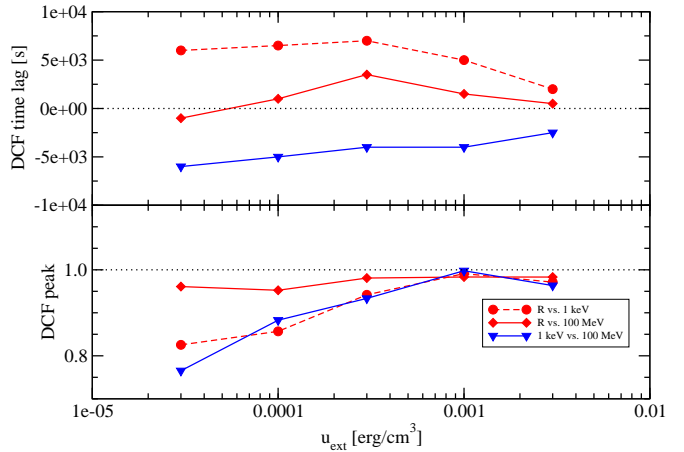}
\caption{\label{cs_dcfpar} Dependence on inter-band cross correlations
and time lags on the external radiation energy density. }
\end{figure}

From the internal-shock simulations, energy-dependent light curves
could be extracted. Using the standard Discrete Correlation Function
(DCF: \cite{ek88}) analysis, inter-band time lags could be extracted
for any set of parameters. Fig. \ref{cs_dcfpar} shows the dependence
of the inter-band time lags between optical, X-ray and \fermi\
$\gamma$-ray light curves (top panel) and the quality of the cross
correlation, characterized by the peak value of the DCF (bottom panel)
as a function of the external radiation energy density.

One of the most remarkable results of this study was that only slight
changes in physical parameters can lead to substantial changes of the
inter-band time lags and even a reversal of the sign of the lags. This
may explain the lack of consistency of lags even within the same source.

\begin{acknowledgements}
I thank Manasvita Joshi for careful proofreading of the manuscript and
helpful comments. I acknowledge support from NASA through Fermi Guest 
Investigator Grants NNX09AT81G and NNX09AT82G, XMM-Newton Guest Investigator 
grant NNX09AV45G, as well as Astrophysics Theory Program grant NNX10AC79G.
\end{acknowledgements}

\end{document}